\documentclass{article}
\usepackage{amsmath,graphicx,mlspconf}
\usepackage{amsfonts}
\usepackage{bm}
\usepackage{multirow}
\usepackage{hyperref}
\usepackage{cite}

\newcommand{\bb}[1]{\bm{\mathrm{#1}}}

\copyrightnotice{978-1-7281-6338-3/21/\$31.00 {\copyright}2021 IEEE}

\toappear{2021 IEEE International Workshop on Machine Learning for Signal Processing, Oct.\ 25--28, 2021, Gold Coast, Australia}

\title{Joint optimization of system design and reconstruction in MIMO radar imaging}
\name{%
    Tomer Weiss%
    \qquad Nissim Peretz %
    \qquad Sanketh Vedula%
    \qquad Arie Feuer%
    \qquad Alex Bronstein%
}
\address{%
    Technion -- Israel Institute of Technology, Haifa, Israel%
}

\begin{document}

\maketitle
\vspace{-0.5cm}
\begin{abstract}
\vspace{-0.1cm}
Multiple-input multiple-output (MIMO) radar is one of the leading depth sensing modalities. However, the usage of multiple receive channels lead to relative high costs and prevent the penetration of MIMOs in many areas such as the automotive industry. Over the last years, few studies concentrated on designing reduced measurement schemes and image reconstruction schemes for MIMO radars, however these problems have been so far addressed separately. On the other hand, recent works in optical computational imaging have demonstrated growing success of simultaneous learning-based design of the acquisition and reconstruction schemes, manifesting significant improvement in the reconstruction quality. 
Inspired by these successes, in this work, we propose to learn MIMO acquisition parameters in the form of receive (Rx) antenna elements locations jointly with an image neural-network based reconstruction. 
To this end, we propose an algorithm for training the combined acquisition-reconstruction pipeline end-to-end in a differentiable way. 
We demonstrate the significance of using our learned acquisition parameters with and without the neural-network reconstruction.\\
Code and datasets will be released upon publication.
\end{abstract}
\vspace{-0.5cm}
\section{Introduction}
\label{sec:intro}
\vspace{-0.3cm}

Imaging technologies play a vital role in the emerging autonomous vehicle ecosystem. There is a wide consensus in this industry that a combination of several long-range (over 100m) depth sensing modalities is imperative for the viability of self-driving cars. RF sensors complement the optical modality in autonomous vehicles. Specifically, millimeter wave multiple-input multiple-output (MIMO) radar \cite{MIMO} is a mature technology providing accurate range, velocity and direction of arrival (DOA) estimation at relatively long distances. MIMO radar can penetrate much denser fog and rain compared to the optical counterparts.

The current weakness of this technology is that to achieve sufficient angular resolution, multiple receive channels are required. This requirement currently dictates the high cost of the device. The ability to maintain high resolution and quality images using a smaller number of receive channels, will significantly reduce these technology costs and increase the commercial viability of automotive digital MIMO radars.
Standard radar processing samples and processes the received signal at its Nyquist rate. However, recent research showing that by using compressed sensing (CS) \cite{donoho2006compressed} even sub-Nyquist sample rate is sufficient for reconstructing the underlying signal. CS was used in MIMO radars to increase resolution \cite{strohmer2013sparse, strohmer2014analysis}, reduce processing time \cite{yu2010mimo_cs} and to reduce the number of antennas used \cite{rossi2013mimo_spatial}.

Recently, machine learning (ML) based approaches were proposed to improve radar signal reconstruction: for example, \cite{wang2018sar_cnn} and \cite{guan2019mimo_cnn} used convolutional neural networks (CNNs) to generate ``visible" images from synthetic-aperture radar (SAR) and MIMO radars respectively. These works collected paired datasets of radar signals and visible images; they train CNNs to ``translate" radar images to visible ones. 
ML techniques were also employed to choose sparse subarray of sensors in the context of cognitive radars \cite{elbir2019sparse_cognitive1, elbir2020sparse_cognitive2}.

\begin{figure*}[!t]
	\centering
	\vspace{-1.3cm}
	\includegraphics[width=0.8\textwidth]{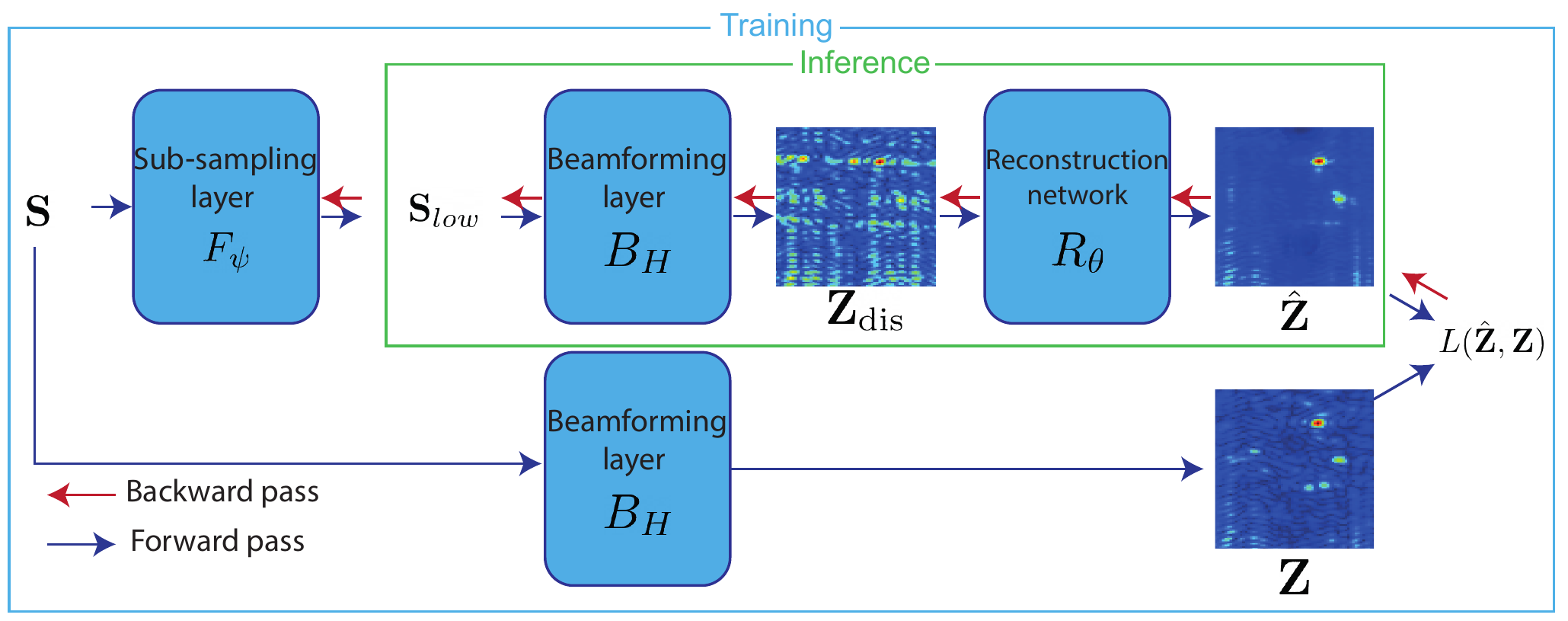} 
	\vspace{-0.3cm}
	\caption{\small{\textbf{Data-flow pipeline for the proposed algorithm.} During training, the forward model parametrized by $\bb{\psi}$ (antenna locations) is optimized together with reconstruction model parametrized by $\bb{\theta}$. At inference,  the optimized antenna locations ($\bb{\psi}$) are fixed and programmed into the hardware to be used during acquisition.}}
	\label{fig:pipeline}
\vspace{-0.5cm}
\end{figure*} 

CS methods use random strategies to select system design parameters. The question remains, ``is it indeed the optimal strategy?”. Recent developments in computational imaging suggest that learned system designs significantly improve the end performance \cite{haim2018depth, vedula2018learning, weiss2019pilot, weiss2020towards, weiss2020joint}. The main purpose of our paper is to exemplify how joint optimization of system design parameters and image reconstruction lead to better performance when compared to random system design strategies. We demonstrate a proof-of-concept of this approach to design optimized location of Rx antennas in MIMO radars.

In our experiments, we compare our learned designs of antenna location design with random designs as used in \cite{rossi2013mimo_spatial}; we report an improvement of $0.88-1.88$dB in PSNR. In addition to learned system design, we propose a CNN-based reconstruction for radar signal reconstruction. Our learned designs, together with CNN-based reconstruction, observe an overall improvement of $4.93-6.36$dB. In particular, we introduce and tackle the following two flavors of the Rx antenna location design problem:
\begin{enumerate}
    \item \bb{Learning discrete locations.} Given an existing MIMO device, the goal is to learn the best {subset} of Rx antennas for a given environment/end-task. This is equivalent to antenna \textit{selection}; the problem is therefore discrete in nature.
    \item \bb{Learning continuous locations.} Instead of focusing on selecting optimal locations within a given MIMO device, the goal here is to learn {new} antennas' location which can be used to design a new device that is optimized given (i) a budget of Rx antennas, (ii) an environment (e.g., indoor, outdoor), and (iii) an end-task of interest (e.g., detection, reconstruction).
\end{enumerate}

\vspace{-0.5cm}
\section{Methods}
\label{sec:methods}
\vspace{-0.3cm}

It is convenient to view our approach as a single neural network combining the forward (acquisition) and the inverse (reconstruction) models (see Fig. \ref{fig:pipeline} for a schematic depiction). During training, the input to the model is a complex baseband signal obtained from the full set of Rx antennas, denoted as $\bb{S}\in \mathbb{C}^{N_\text{range} \times N_{\text{virt}}}$ where $N_\text{range}$ and $N_{\text{virt}}$ are the number of range bins and number of virtual elements, respectively. In MIMO radars, $N_{\text{virt}} = N_TN_R$, where $N_T$ and $N_R$ are, respectively, the number of transmit and receive antennas in the full set of antennas \cite{MIMO}. The input is faced by a sub-sampling layer modeling the data acquisition at the $n_R<N_R$ learned Rx antennas locations, a beamforming layer producing a range-azimuth map, and an end-task model operating in the range-azimuth domain and producing a reconstructed range-azimuth map (i.e., signal reconstruction) or any other downstream task (e.g., detection, localization, segmentation).

 In what follows, we detail the three ingredients of our pipeline. It should be mentioned that all components are differentiable with respect to the $n_R$ antennas locations, denoted collectively as $\bb{\psi}$, in order to allow training the latter with respect to the performance of the end-task of interest.

\vspace{-0.3cm}
\subsection{Sub-sampling layer}
\label{ssec:subsampling}
\vspace{-0.15cm}
The goal of the sub-sampling layer, denoted by $F_{\bb{\psi}}: \\ \mathbb{C}^{N_\text{range} \times N_TN_R} \to \mathbb{C}^{N_\text{range} \times N_Tn_R}$, is to create the set of measurements to be acquired by each one of the Rx antennas according to their locations $\bb{\psi}$. 
Both in \textit{discrete selection} and \textit{continuous sampling} the output $\bb{S}_\text{low} = F_{\bb{\psi}}(\bb{S}) \in \mathbb{C}^{N_\text{range} \times N_Tn_R}$ is of the same functionality (i.e., both emulate the sub-sampled signal given a fully-sampled one), but the algorithms to emulate it are different. In the following subsections, we describe the algorithms to emulate $\bb{S}_\text{low}$ from $\bb{S}$ and $\bb{\psi}$ in these two scenarios.\footnote{In the interest of maintaining continuity of the manuscript and delivering the message of the paper effectively, the details of how discrete selection and continuous sampling are performed are deferred to Appendix A and B, respectively.}
\vspace{-0.35cm}
\subsubsection{Discrete selection}
\label{sssec:§}
\vspace{-0.15cm}
In the discrete selection scenario, $\bb{\psi}$ is parametrized as a vector of size $N_R$. Each entry in this vector is the probability to select the corresponding Rx antenna. First, we tried using the Gumbel-softmax reparametrization technique \cite{jang2016softmax} to learn the weights $\bb{\psi}$. However, we noticed that using this trick off-the-shelf would mean that all the channels are treated independently, meaning that the selection of an antenna does not affect the probability of selecting nearby antennas. This is unrealistic as the signals received at neighboring antennas are likely to contain similar information.

To overcome this drawback, inspired by \cite{wang2020mv_bernoulli}, we use the multi-variate analogue of the Gumbel-softmax trick that learns also a covariance parameter to represent the correlation between the selection probability of the different antennas. For further details, we refer the reader to Appendix A; note that at inference time, the antennas with the top $n_R$ weights are selected.
\vspace{-0.35cm}
\subsubsection{Continuous sampling}
\label{sssec:continus}
\vspace{-0.15cm}
In the continuous learning scenario $\bb{\psi}$ is parametrized as a vector of size $n_R$. Each entry in this vector is the coordinate of an antenna, where the distance between two consecutive antennas is the unit of measure (for e.g. $\bb{\psi}_2 = 3.5$ means that the location of the 2nd antenna is exactly between the 3rd and 4th antennas of the original device).

Emulating the signal of this new location can be achieved via linearly interpolating the signals received at the two closest antennas. We observed that linear interpolation reduces the noise differently as function of the learned coordinate, leading to false local minima between every two antennas; it causes the optimization to get stuck in these regions. To overcome this issue, we added a second acquisition of the same scene and interpolated between the two temporally consecutive acquisitions in such a way that will lead to a constant noise reduction as function of the learned coordinate. For further details, we refer the reader to Appendix B.
\vspace{-0.4cm}
\subsection{Beamforming layer}
\vspace{-0.15cm}
For transforming the raw signal $\bb{S}_\text{low}$ to the Range-Azimuth domain, we perform standard delay-and-sum beamforming using a predefined steering matrix $\bb{H}$, henceforth denoted by $B_{\bb{H}}$. The steering matrix represents the set of phase-delays at each virtual antenna element. The beamforming is performed by multiplying $\bb{S}_\text{low}$ and $\bb{H}$ taking the mean over the virtual elements dimension followed by an inverse FFT along the range dimension and outputs the magnitude of the result. The result is a (distorted) Range-Azimuth map, $\bb{Z}_\textrm{dis} = B_{\bb{H}} (\bb{S}_\text{low})$. 
\vspace{-0.4cm}
\subsection{Task model}
\vspace{-0.15cm}
The goal of the task model is to extract the representation of the distorted Range-Azimuth map $\bb{Z}_\textrm{dis}$ that will contribute the most to the performance of the end-task such as reconstruction, localization or segmentation. At training, the task-specific performance is quantified by a loss function, which is described in Section \ref{subsec:loss}.  
The model is henceforth denoted by $\hat{\bb{Z}} = R_{\bb{\theta}}(\bb{Z}_\textrm{dis})$, with $\bb{\theta}$ representing its learnable parameters. The input to the network is the distorted Range-Azimuth map, $\bb{Z}_\textrm{dis}$, while the output varies according to the specific task. For example, in reconstruction, the output is a Range-Azimuth map, while in segmentation it is a mask representing the segments of the observed scene.

In the present work, to implement the task model we used a multi-resolution encoder-decoder network with symmetric skip connections, also known as the U-Net architecture \cite{ronneberger2015unet}. U-Net has been widely-used in computer vision tasks such as reconstruction \cite {zbontar2018fastmri} and segmentation \cite{ronneberger2015unet} and also in radar applications \cite{radar_unet_1, radar_unet_2}. It is important to emphasize that the principal focus of this work is not on the task model \emph{in se}, since the proposed algorithm can be used with any differentiable model to improve the end task performance.
\vspace{-0.4cm}
\subsection{Loss function and training}
\label{subsec:loss}
\vspace{-0.15cm}

The training of the proposed pipeline is performed by simultaneously learning the antennas locations $\bb{\psi}$ and the parameters of the task model $\bb{\theta}$.

The aim of the loss is to measure how well the specific end task is performed.  For the reconstruction task we performed in this work, we chose the $L_2$ norm to measure the discrepancy between the model output image $\hat{\bb{Z}}$ and the ground-truth image $\bb{Z}=B_{\bb{H}} (\bb{S})$, derived by beamforming the measurement acquired with the full set of Rx antennas. In this case, the loss is $L = \| \hat{\bb{Z}} - \bb{Z} \|_2$. Similarly, for any other end-task (e.g., detection, tracking), we can measure the discrepancy between $\hat{\bb{Z}}$ and the relevant groundtruth.

The training is performed by solving the optimization problem 
\begin{equation}
\min_{\bb{\psi}, \bb{\theta}} \, \sum_{\bb{S}, \bb{Z}} \ell( R_{\bb{\theta}}(B_{\bb{H}}(F_{\bb{\psi}} (\bb{S}) ) ) , \bb{Z}),
\label{eq:min}
\end{equation}
where the loss is summed over a training set comprising the pairs of fully sampled data $\bb{S}$ and the corresponding groundtruth output $\bb{Z}$.
\vspace{-0.3cm}
\section{EXPERIMENTS AND DISCUSSION}
\label{sec:expriments}
\vspace{-0.3cm}

\subsection{Experimental settings}
\vspace{-0.15cm}
In our experiments, we used a dataset we acquired using IMAGEVK-74\footnote{\url{https://www.minicircuits.com/WebStore/vtrig_74.html}} 4D millimeter wave imaging kit.
The kit includes 20 Tx and 20 Rx on-board antennas that can be configured to transmit and receive signals anywhere within the 62 to 69 GHz range. We used the full set of antennas $N_T=N_R=20$ and $N_\text{range}=75$ range bins. The dataset comprises metal objects randomly positioned in various DOA angles, received by a uniform linear virtual array of 400 virtual elements.
Our training and test sets comprised  $2700$ and $300$ acquisitions, respectively. The reconstructed Range-Azimuth map was compared to the Range-Azimuth map obtained using the full set of channels. For quantitative evaluation, we used the peak signal-to-noise ratio (PSNR) and structural-similarity measure (SSIM). In all our experiments, we compared between learned and random Rx antennas locations over the entire test set\footnote{To enable fair comparison, both in the discrete selection and continuous sampling cases we used the same noise reduction, see Appendix B.}. For the random selections, we evaluated $10$ random selections and use the one with the best performance.
In all experiments, we used the Adam \cite{kingma2017adam} optimizer with learning rate of $0.001$ and $200$ epochs.
\vspace{-0.4cm}
\subsection{Results}
\vspace{-0.15cm}
Both in the discrete selection and continuous sampling scenarios, we observed superior performance of the learned Rx antenna locations. Table \ref{table:discrete} summarizes the quantitative results for different amount of Rx channels for the discrete scenario and Table \ref{table:contious} for the continuous scenario. The final image quality is affected by two factors:
\begin{enumerate}
    \vspace{-0.1cm}
    \item \textbf{Neural network reconstruction} leads to significant improvement of $2.16-5.53$dB in PSNR and $0.177-0.435$ SSIM points.
    \vspace{-0.1cm}
    \item \textbf{Antenna location learning} leads to an additional improvement of $0.32-2.58$dB in PSNR and $0.039-0.282$ SSIM points without reconstruction and $0.88-1.88$dB in PSNR and $0.061-0.128$ SSIM points with reconstruction.
    \vspace{-0.1cm}
\end{enumerate}
The total improvement is $4.93-6.36$dB in PSNR and $0.364-0.512$ SSIM points.
Although the discrete optimization domain is a subset of the continuous domain, we can see that in some cases the learned discrete locations produces better results than the learned continuous locations. This is however true only before reconstruction. The loss was applied to the post-reconstruction Range-Azimuth maps, with the learned continuous locations leading to better (or comparable) results.

\begin{table*}
\vspace{-0.2cm}
    \centering
\begin{tabular}{|c|c|c|c|c|c|}
\hline
\multirow{2}{*}{$n_R$}&\multirow{2}{*}{Antennas' locations}&\multicolumn{2}{|c|}{Without reconstruction}&\multicolumn{2}{|c|}{With reconstruction}\\
\cline{3-6}
&&PSNR&SSIM&PSNR&SSIM\\
\hline
\multirow{2}{*}{5}&Random&22.68$\pm$0.67&0.184$\pm$0.014&27.16$\pm$0.50&0.450$\pm$0.017\\
&Learned&\textbf{25.26$\pm$0.77}&\textbf{0.332$\pm$0.017}&\textbf{28.39$\pm$0.67}&\textbf{0.578$\pm$0.015}\\
\hline
\multirow{2}{*}{7}&Random&23.77$\pm$0.74&0.191$\pm$0.017&28.73$\pm$0.58&0.579$\pm$0.021\\
&Learned&\textbf{25.84$\pm$0.79}&\textbf{0.339$\pm$0.023}&\textbf{\textbf{29.63$\pm$0.53}}&\textbf{0.640$\pm$0.013}\\
\hline
\multirow{2}{*}{10}&Random&24.87$\pm$0.75&0.211$\pm$0.023&29.85$\pm$0.60&0.619$\pm$0.018\\
&Learned&\textbf{27.67$\pm$0.78}&\textbf{0.493$\pm$0.023}&\textbf{31.23$\pm$0.65}&\textbf{0.723$\pm$0.012}\\
\hline
\end{tabular}
\vspace{-0.3cm}
\caption{\small{\textbf{Comparison of random and learned antennas locations in the discrete selection scenario.} Presented are the PSNR and SSIM metrics of the random and learned antennas locations for different numbers of Rx antennas (\textrm{$n_R$}) with and without reconstruction.}}
\label{table:discrete}
\end{table*} 

\begin{table*}
    \centering
\vspace{-0.2cm}
\begin{tabular}{|c|c|c|c|c|c|}
\hline
\multirow{2}{*}{$n_R$}&\multirow{2}{*}{Antennas' locations}&\multicolumn{2}{|c|}{Without reconstruction}&\multicolumn{2}{|c|}{With reconstruction}\\
\cline{3-6}
&&PSNR&SSIM&PSNR&SSIM\\
\hline
\multirow{2}{*}{5}&Random&22.57$\pm$0.71&0.185$\pm$0.014&27.32$\pm$0.61&0.507$\pm$0.017\\
&Uniform&22.20$\pm$0.63&0.161$\pm$0.017&27.73$\pm$0.71&0.494$\pm$0.021\\
&Learned&\textbf{23.25$\pm$0.61}&\textbf{0.178$\pm$0.013}&\textbf{28.61$\pm$0.80}&\textbf{0.584$\pm$0.018}\\
\hline
\multirow{2}{*}{7}&Random&23.75$\pm$0.66&0.179$\pm$0.015&28.40$\pm$0.91&0.575$\pm$0.022\\
&Uniform&23.48$\pm$0.63&0.183$\pm$0.013&28.14$\pm$0.88&0.532$\pm$0.028\\
&Learned&\textbf{24.07$\pm$0.62}&\textbf{0.212$\pm$0.012}&\textbf{\textbf{29.70$\pm$0.74}}&\textbf{0.647$\pm$0.019}\\
\hline
\multirow{2}{*}{10}&Random&26.24$\pm$0.72&0.378$\pm$0.018&29.29$\pm$0.90&0.631$\pm$0.020\\
&Uniform&26.83$\pm$0.85&0.445$\pm$0.023&28.99$\pm$0.71&0.622$\pm$0.016\\
&Learned&\textbf{27.86$\pm$0.83}&\textbf{0.469$\pm$0.023}&\textbf{31.17$\pm$0.69}&\textbf{0.747$\pm$0.015}\\
\hline
\end{tabular}
\vspace{-0.2cm}
\caption{\small{\textbf{Comparison of random and learned antennas locations in the continuous sampling scenario.} Presented are the PSNR and SSIM metrics of the random, uniform and learned antennas locations for different numbers of Rx antennas (\textrm{$n_R$}) with and without reconstruction.}}
\label{table:contious}
\vspace{-0.3cm}
\end{table*} 

\begin{figure}
\vspace{-0.2cm}
	\centering
	\includegraphics[width=1\linewidth]{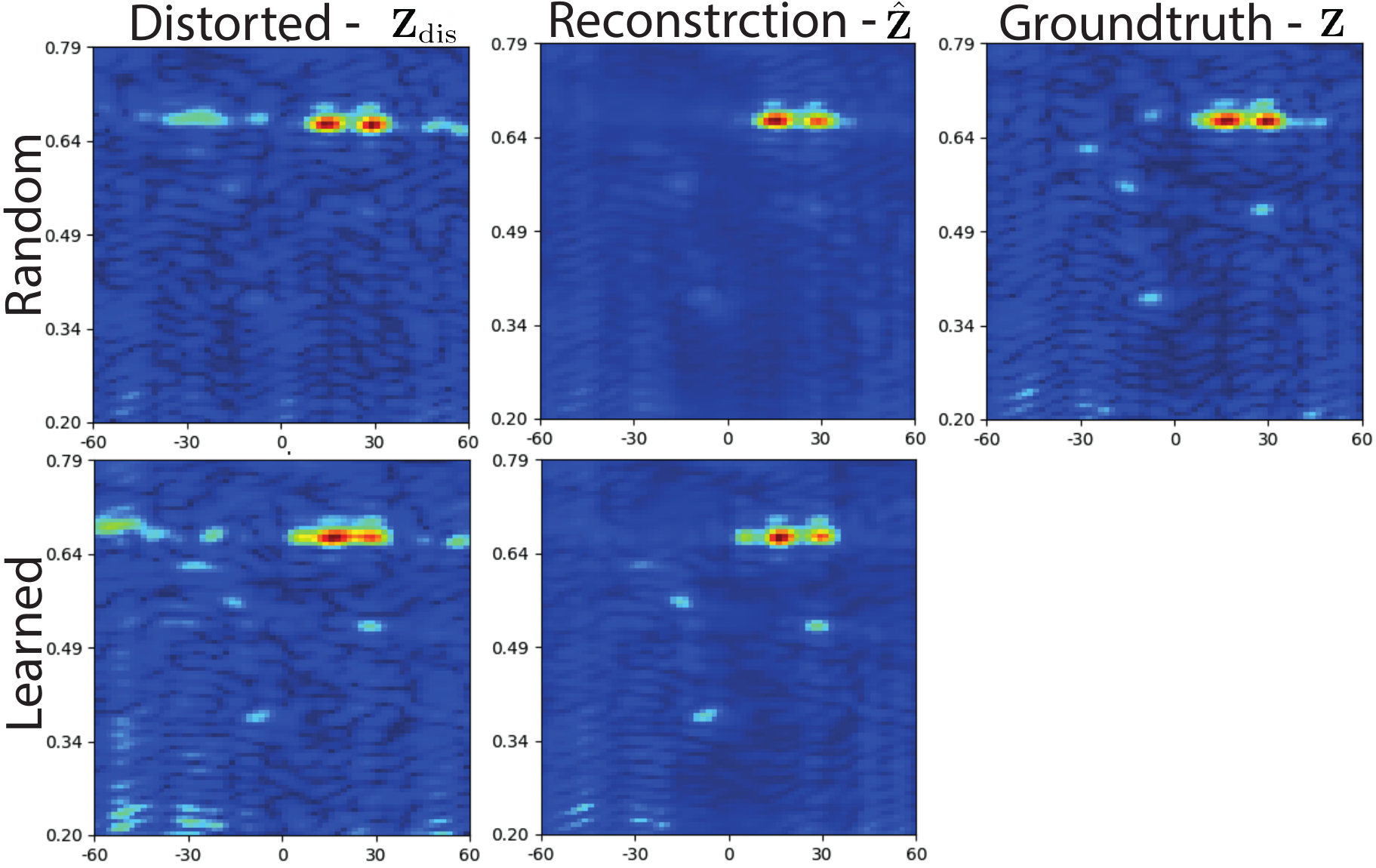} 
	\vspace{-0.8cm}
	\caption{\small{\textbf{Visual comparison.} Comparison between the Range-Azimuth maps of the random and learned channel selection in the discrete learning scenario using budget of $n_R=10$ Rx channels.}}
	\label{fig:visual}
\vspace{-0.62cm}
\end{figure}

Visual results depicted in Figure \ref{fig:visual} clearly demonstrate the importance of the learned channel selection. We can clearly locate two strong reflectors and 5 weaker reflectors in this scene.
The learned antenna selection contains the most important data for the reconstruction network to be able to reconstruct the best Range-Azimuth map. As we assume, the reconstruction of the learned selection was able to locate more targets than the random selection. 

\begin{figure}[t]
	\centering
	\includegraphics[width=1\linewidth]{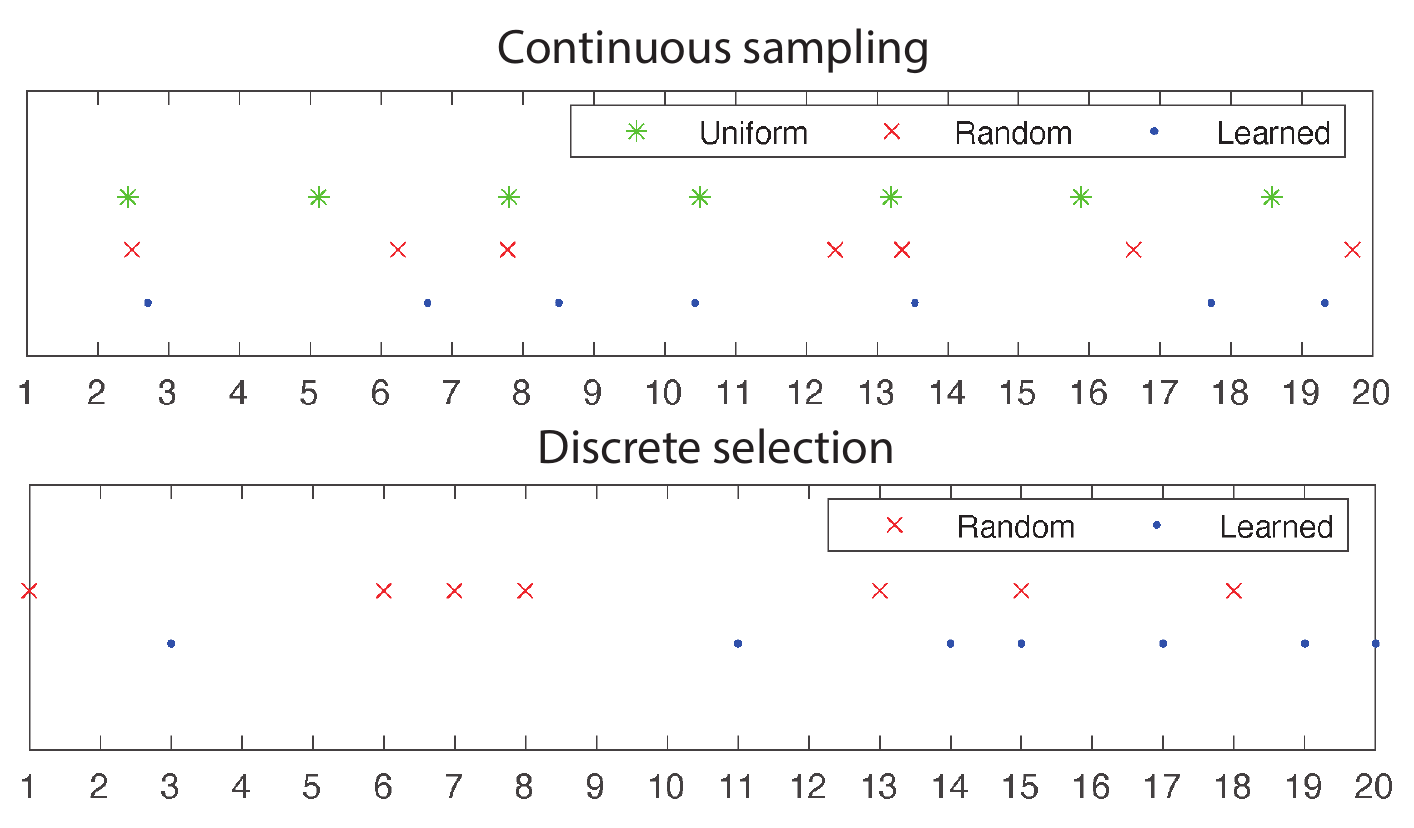} 
	\vspace{-0.8cm}
	\caption{\small{\textbf{Antenna locations.} Comparison between random, uniform and learned antenna locations in the discrete selection and continuous sampling scenarios.}}
	\label{fig:locations}
\vspace{-0.6cm}
\end{figure}

Another important benefit of the learned selection compared to a random selection is its multivariate nature. The learned selection is the best combination of antennas, each antenna is chosen to deliver the best additive performance with respect to all other antenna locations. This is in contrast to random selection in which each antenna is chosen randomly. Comparison between random, uniform and learned antenna locations is presented in Fig. \ref{fig:locations} 

\vspace{-0.3cm}
\section{CONCLUSION AND FUTURE DIRECTIONS}
\label{sec:conclusion}
\vspace{-0.3cm}

We have demonstrated, as a proof-of-concept, that learning the Rx antenna locations simultaneously with a reconstruction neural model improves the end quality of a MIMO radar. The quality improvement in this work arises from two factors which can be used independently or in concert. The first factor is the neural-network reconstruction, and its training scheme, which leads to significant improvement in the image quality. The second factor is the acquisition parameter optimization in the form of Rx antenna locations which lead to additive improvement. In cases where neural network based reconstruction is not desired, one can still use the learned acquisition parameter without the fear of neural network hallucination or interpolation.

In this work, we showed how to learn Rx antenna locations together with Range-Azimuth map reconstruction. In future work, we plan to explore and optimize additional acquisition parameters such as the transmitted waveform modulation, Doppler shift acquisition, and more. We also plan to explore more downstream tasks such as detection, localization and segmentation.
\vspace{-0.4cm}
\section{APPENDIX A: The Multivariate Gumbel-softmax trick}
\label{sec:appendix_a}
\vspace{-0.3cm}
The standard Gumbel-softmax reparametrization technique \cite{jang2016softmax} allows to sample from discrete random variables in a differentiable manner. A similar technique allows to sample from a relaxed Bernoulli distribution, \\$B \sim \text{RelaxedBernoulli}(\alpha, \lambda)$, as follows:
$$U \sim \text{Uniform}(0, 1),$$
$$L = \log(\alpha) + \log(U) - \log(1 - U),$$
$$B = \frac{1}{1 + \exp(-L/\lambda)},$$
where $\alpha$ is the location parameter and $\lambda$ is the temperature parameter that controls the degree of approximation. \\
We were inspired by \cite{wang2020mv_bernoulli} who proposed to use the Gaussian copula to characterize the correlation between multiple uniform random variables, so that their dependencies can be transferred to multiple relaxed Bernoulli variables. We added a learned covariance $\bb{\Sigma}$ to the sampling process. In order to keep $\bb{\Sigma}$ positive semi-definite, we used its factor $\bb{L}$ as the learned parameter. In order to sample $\bb{\psi}$ from this new Top K distribution ($\bb{\psi} \sim \text{TopK}()$), we used the following algorithm:
\begin{enumerate}
\itemsep0em
    \item Draw a standard normal sample: $\bb{\epsilon} \sim N (\bb{0}, \bb{I})$
    \item Generate a multivariate Gaussian vector: $g = \bb{L}\bb{\epsilon}$
    \item Apply element-wise Gaussian CDF $\Phi_{\sigma_i}$ with mean zero and variance $\sigma_i^2$, where $\sigma_i=\Sigma_{ii}=(\bb{L}\bb{L}^T)_{ii}$: $$U_i = \Phi_{\sigma_i}(g_i)$$
    \item Apply inverse CDF of the logistic distribution: $$l_i = 
    \log(\alpha_i) + \log(U_i) - \log(1 - U_i),$$
    \item Apply relaxed Top K operator \cite{xie2019topk}: $$\bb{\psi}=\text{RelaxedTopK}(l, n, \lambda)$$
\end{enumerate}
In all of our experiments we used $\lambda=0.001$.
Using this reparametrization, we need to replace the optimization variables in (\ref{eq:min}), obtaining the following problem.
$$\min_{\bb{\alpha}, \bb{L}, \bb{\theta}} \, \sum_{\bb{S}, \bb{Z}} \ell( R_{\bb{\theta}}(B_{\bb{H}}(F_{\bb{\psi}} (\bb{S}) ) ) , \bb{Z})$$
\vspace{-0.8cm}
\section{APPENDIX B: Emulating continuous antenna location sampling}
\label{sec:appendix_b}
\vspace{-0.3cm}
We observed that when creating a synthetic signal that would be acquired by an antenna between two actually sampled antennas, linear interpolation with weights set according to the antenna's relative position affects the SNR in a location-dependent way.

\begin{figure}[!h]
	\centering
	\includegraphics[width=0.9\linewidth]{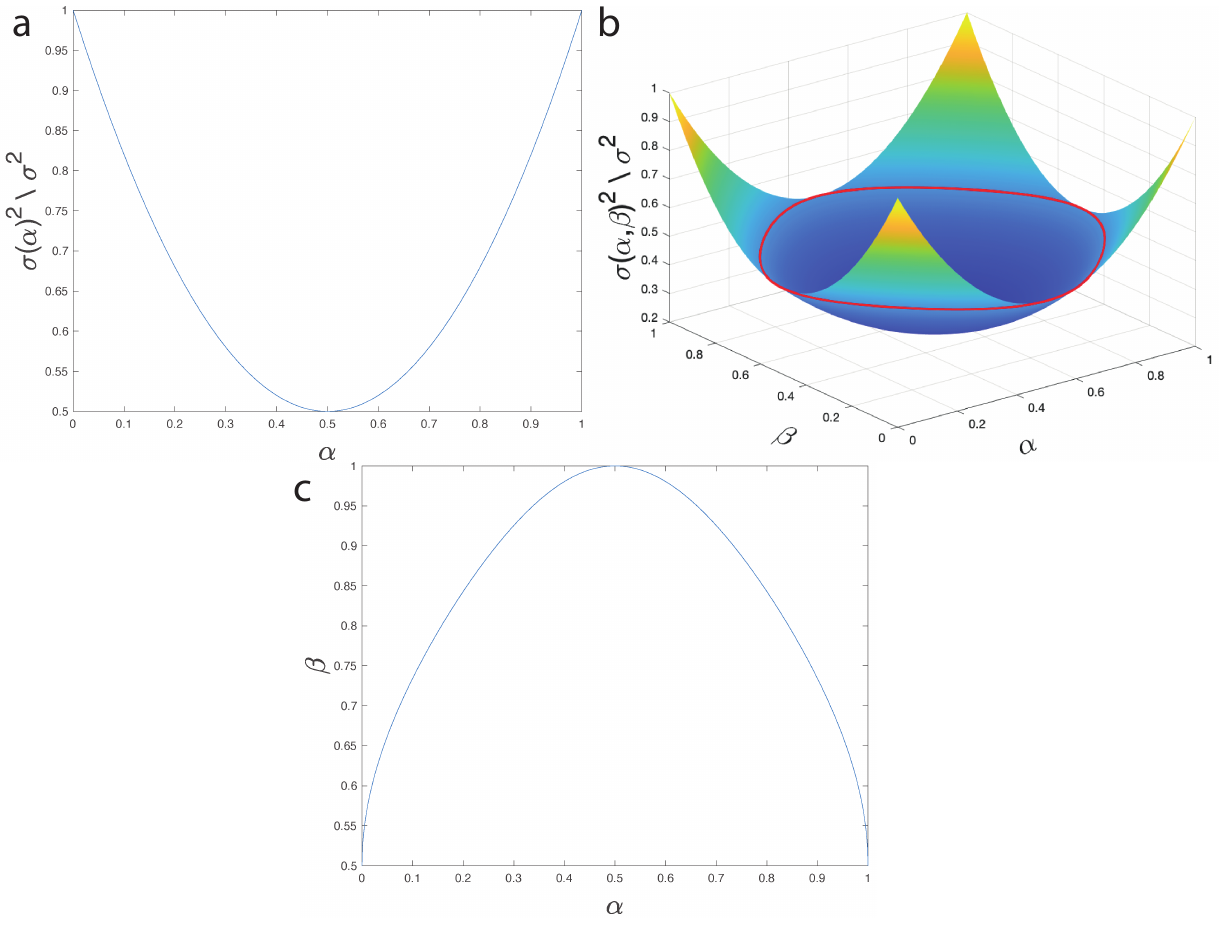} 
	\vspace{-0.6cm}
	\caption{\small{(a) Noise reduction as function of the interpolation weight in 2 data channel interpolation. (b) Noise reduction as function of the interpolation weight in 4 data channel interpolation. Red line represents the noise level when using the function in (c) to determine $\beta$ according to $\alpha$.}}
	\label{fig:noise}
\vspace{-0.5cm}
\end{figure}

Assume that two antennas located at positions $i$ and $i+1$ receive the signals $c_i$ and $c_{i+1}$, respectively; the signal received at a new antenna placed at $i+\alpha$ can be emulated as the linear combination $c(\alpha)=(1-\alpha) c_i + \alpha c_{i+1}$. 
Assuming both channels contain independent Gaussian noise with variance $\sigma^2$, the noise level of $c(\alpha)$ is $\sigma(\alpha)^2= ((1-\alpha)\sigma)^2 + (\alpha \sigma)^2=((1-\alpha)^2 + \alpha^2)\sigma^2$, as depicted in Fig. \ref{fig:noise}a.

When optimizing a loss that depends on $c(\alpha)$ with respect to $\alpha$, false local minima arise in the optimization landscape that are related only to our emulation and not to the original problem. To overcome this issue, we added a second acquisition of the same scene and interpolated between the two consecutive acquisitions in such a way that will lead to a constant noise reduction as function of the learned coordinate. 
Specifically, we assume the previous two channels to be sampled twice -- at time $t$ and $t+1$, yielding four realizations $c_{t,i}, c_{t,i+1}, c_{t+1,i}$ and $c_{t+1,i+1}$. We introduce a new weight $\beta$ for the bilinear interpolation in space and time, $c(\alpha, \beta)=(1 - \beta)((1-\alpha) c_{t,i} + \alpha c_{t,i+1})+\beta((1-\alpha) c_{t+1,i} + \alpha c_{t+1,i+1})$. The noise of variance of the new synthetic channel now becomes $\sigma(\alpha, \beta)^2=((1-\beta)^2+\beta^2)((1-\alpha)^2+\alpha^2)\sigma^2$
as depicted in Fig. \ref{fig:noise}b. Setting $\beta=\frac{1}{2}(1+\sqrt{-1+\frac{1}{2\alpha^2-2\alpha+1}})$ (Fig. \ref{fig:noise}c) yields a constant SNR function $\sigma(\alpha)^2=0.5\sigma^2$, for any choice of $\alpha$ (red line in Fig. \ref{fig:noise}b). 

We believe that this simple technique is useful in other applications where the need to synthetically create a new channel by weighted interpolation of recorded channels arises. 

\bibliographystyle{IEEEbib}
\vspace{-0.4cm}
\bibliography{refs}

\begin{thebibliography}{10}

\bibitem{MIMO}
E.~Fishler, A.~Haimovich, R.~Blum, D.~Chizhik, L.~Cimini, and R.~Valenzuela,
\newblock ``Mimo radar: an idea whose time has come,''
\newblock in {\em Proc. of the IEEE Radar Conference (RadarConf)}, 2004.

\bibitem{donoho2006compressed}
David~L Donoho,
\newblock ``Compressed sensing,''
\newblock {\em IEEE Transactions on information theory}, vol. 52, no. 4, pp.
  1289--1306, 2006.

\bibitem{strohmer2013sparse}
Thomas Strohmer and Haichao Wang,
\newblock ``Sparse mimo radar with random sensor arrays and kerdock codes,''
\newblock in {\em Proc. IEEE Int. Conf. Sampling Theory and Applications},
  2013, pp. 517--520.

\bibitem{strohmer2014analysis}
Thomas Strohmer and Benjamin Friedlander,
\newblock ``Analysis of sparse mimo radar,''
\newblock {\em Applied and Computational Harmonic Analysis}, vol. 37, no. 3,
  pp. 361--388, 2014.

\bibitem{yu2010mimo_cs}
Yao Yu, Athina~P Petropulu, and H~Vincent Poor,
\newblock ``Mimo radar using compressive sampling,''
\newblock {\em IEEE Journal of Selected Topics in Signal Processing}, vol. 4,
  no. 1, pp. 146--163, 2010.

\bibitem{rossi2013mimo_spatial}
Marco Rossi, Alexander~M Haimovich, and Yonina~C Eldar,
\newblock ``Spatial compressive sensing for mimo radar,''
\newblock {\em IEEE Transactions on Signal Processing}, vol. 62, no. 2, pp.
  419--430, 2013.

\bibitem{wang2018sar_cnn}
Puyang Wang and Vishal~M Patel,
\newblock ``Generating high quality visible images from sar images using
  cnns,''
\newblock in {\em IEEE Radar Conference (RadarConf)}, 2018.

\bibitem{guan2019mimo_cnn}
Junfeng Guan, Sohrab Madani, Suraj Jog, and Haitham Hassanieh,
\newblock ``High resolution millimeter wave imaging for self-driving cars,''
\newblock {\em arXiv preprint arXiv:1912.09579}, 2019.

\bibitem{elbir2019sparse_cognitive1}
Ahmet~M Elbir, Satish Mulleti, Regev Cohen, Rong Fu, and Yonina~C Eldar,
\newblock ``Deep-sparse array cognitive radar,''
\newblock in {\em 2019 13th Int'l Conf. on Sampling Theory and Applications
  (SampTA)}. IEEE, 2019.

\bibitem{elbir2020sparse_cognitive2}
Ahmet~M Elbir and Kumar~Vijay Mishra,
\newblock ``Sparse array selection across arbitrary sensor geometries with deep
  transfer learning,''
\newblock {\em IEEE Transactions on Cognitive Communications and Networking},
  2020.

\bibitem{haim2018depth}
Harel Haim, Shay Elmalem, Raja Giryes, Alex~M Bronstein, and Emanuel Marom,
\newblock ``Depth estimation from a single image using deep learned phase coded
  mask,''
\newblock {\em IEEE Transactions on Computational Imaging}, vol. 4, no. 3, pp.
  298--310, 2018.

\bibitem{vedula2018learning}
Sanketh Vedula, Ortal Senouf, Grigoriy Zurakhov, Alex Bronstein, Oleg
  Michailovich, and Michael Zibulevsky,
\newblock ``Learning beamforming in ultrasound imaging,''
\newblock {\em Proc. Medical Imaging with Deep Learning (MIDL)}, 2019.

\bibitem{weiss2019pilot}
Tomer Weiss, Ortal Senouf, Sanketh Vedula, Oleg Michailovich, Michael
  Zibulevsky, and Alex Bronstein,
\newblock ``Pilot: Physics-informed learned optimized trajectories for
  accelerated mri,''
\newblock {\em Journal of Machine Learning for Biomedical Imaging (MELBA)},
  2021.

\bibitem{weiss2020towards}
Tomer Weiss, Sanketh Vedula, Ortal Senouf, Oleg Michailovich, et~al.,
\newblock ``Towards learned optimal q-space sampling in diffusion mri,''
\newblock {\em Proc. Computational Diffusion MRI, MICCAI}, 2020.

\bibitem{weiss2020joint}
Tomer Weiss, Sanketh Vedula, Ortal Senouf, Oleg Michailovich, Michael
  Zibulevsky, and Alex Bronstein,
\newblock ``Joint learning of cartesian under sampling andre construction for
  accelerated mri,''
\newblock in {\em International Conference on Acoustics, Speech and Signal
  Processing (ICASSP)}. IEEE, 2020.

\bibitem{jang2016softmax}
Eric Jang, Shixiang Gu, and Ben Poole,
\newblock ``Categorical reparameterization with gumbel-softmax,''
\newblock {\em arXiv preprint arXiv:1611.01144}, 2016.

\bibitem{wang2020mv_bernoulli}
Xi~Wang and Junming Yin,
\newblock ``Relaxed multivariate bernoulli distribution and its applications to
  deep generative models,''
\newblock in {\em Conference on Uncertainty in Artificial Intelligence}. PMLR,
  2020, pp. 500--509.

\bibitem{ronneberger2015unet}
Olaf Ronneberger, Philipp Fischer, and Thomas Brox,
\newblock ``U-net: Convolutional networks for biomedical image segmentation,''
\newblock in {\em MICCAI}, 2015.

\bibitem{zbontar2018fastmri}
Jure Zbontar, Florian Knoll, and Anuroop et~al. Sriram,
\newblock ``fast{MRI}: An open dataset and benchmarks for accelerated {MRI},''
\newblock {\em arXiv preprint arXiv:1811.08839}, 2018.

\bibitem{radar_unet_1}
Michael Stephan and Avik Santra,
\newblock ``Radar-based human target detection using deep residual u-net for
  smart home applications,''
\newblock in {\em 18th IEEE International Conference on Machine Learning And
  Applications}, 2019.

\bibitem{radar_unet_2}
Longhao Xie, Qing Zhao, Chunguang Ma, Binbin Liao, and Jianjian Huo,
\newblock ``{Ü-Net: Deep-Learning Schemes for Ground Penetrating Radar Data
  Inversion},''
\newblock {\em Journal of Environmental and Engineering Geophysics}, 2020.

\bibitem{kingma2017adam}
Diederik~P. Kingma and Jimmy Ba,
\newblock ``Adam: A method for stochastic optimization,'' 2017.

\bibitem{xie2019topk}
Sang~Michael Xie and Stefano Ermon,
\newblock ``Reparameterizable subset sampling via continuous relaxations,''
\newblock {\em arXiv preprint arXiv:1901.10517}, 2019.

\end{thebibliography}

\end{document}